\documentclass[%
twocolumn,
longbibliography,
 amsmath,amssymb,
aps,
pra,
]{revtex4-1}
\usepackage{listings}
\usepackage{graphicx}% Include figure files
\usepackage{dcolumn}% Align table columns on decimal point
\usepackage{bm}% bold math
\usepackage{ulem}
\usepackage{natbib}
\usepackage[colorlinks,citecolor=red,linkcolor=blue,urlcolor=blue]{hyperref}
\usepackage{capt-of}
\usepackage{hhline}
\usepackage{float}
\usepackage{mhchem}
\usepackage[nodisplayskipstretch]{setspace}
\usepackage{calrsfs}
	\DeclareMathAlphabet{\pazocal}{OMS}{zplm}{m}{n}
\usepackage{physics}
\usepackage{appendix}
\usepackage[export]{adjustbox}

\newcommand{\bk}{\mathbf{k}}

\newcommand{\squeezeB}[3]{\left\langle #1\middle| #2\middle| #3\right\rangle}

\usepackage[dvipsnames]{xcolor}
\definecolor{pink}{rgb}{0.858, 0.188, 0.478}

\usepackage{lineno,blindtext}

\begin{document}

%\preprint{APS/123-QED}

\title{Discrete Time Crystal Made of Topological Edge Magnons}% Force line breaks with \\

\author{Dhiman Bhowmick$^1$\href{https://orcid.org/0000-0001-7057-1608}{\includegraphics[scale=0.12]{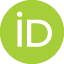}}}
\author{Hao Sun$^2$}
\author{Bo Yang$^1$}
\author{Pinaki Sengupta$^1$\href{https://orcid.org/0000-0003-3312-2760}{\includegraphics[scale=0.12]{orcid.png}}}%
\affiliation{%
 $^1$School of Physical and Mathematical Sciences, Nanyang Technological University, Singapore 
}%
\affiliation{%
 $^2$The Institute for Functional Intelligent Materials(I-FIM),National University of Singapore,4 Science Drive 2,Singapore 117544,Singapore
}%

\date{\today}% It is always \today, today,

\begin{abstract}
We report the emergence of time-crystalline behavior in the $\pi$-Berry phase protected edge states of a Heisenberg ferromagnet in the presence of an external driving field. 
%introduce a time crystal made of the amplified topologically protected edge magnons on a Kagome ferromagnet.
%Analogous to space crystal, time crystal is a phase of periodic events in many-body system which emerges due to spontaneous time-transnational symmetry breaking.
%In this study, we show that in the absence of effective time reversal symmetry breaking Dzyaloshinskii-Moriya interaction, time crystal emerges in $\pi$-Berry phase protected edge state due to spontaneous magnon amplification via external electromagnetic\,(EM) field.
The magnon amplification due to the external field spontaneously breaks the discrete time-translational symmetry, resulting in a discrete time crystal with a period that is twice that of the applied EM field. 
We discuss the nature and symmetry protection of the time crystalline edge states and their stability against various perturbations that are expected in real quantum magnets.
We propose an experimental signature to unambiguously detect the time crystalline behavior and identify two recently discovered quasi-2D magnets as potential hosts. 
We present a first-of-its-kind realization of time crystals at topological edge states, which can be generalized and extrapolated to other bosonic quasi-particle systems that exhibit parametric pumping and topological edge states.
%stability of the state in the parameter space of  
%The measurement of oscillation of transverse spin components at the edge reveal the time crystal phase.
%We identify the regime of stability of the time crystal. 
%We propose that high 
%magnetic field, low temperature and amplification strength.
%can stabilize the time crystal.
\end{abstract}

\pacs{Valid PACS appear here}% PACS, the Physics and Astronomy
\maketitle

%\tableofcontents

%\section{\label{sec1}Introduction}
\section{\label{Section::I}Introduction}
Symmetries and symmetry breaking underlie many interesting phases and phenomena in condensed matter physics. A crystal with a periodic array of atoms/molecules is a simple example where continuous symmetry in space is spontaneously broken. Based on  Lorentz invariance that puts spatial and temporal coordinates on equal footing, Wilczek in 2012 proposed the idea of a time crystal\,\cite{Wilczek}, where time translation symmetry can also be spontaneously broken in the ground state of a quantum many body system -- local observables oscillate in time with fixed periodicity, analogous to the spatial modulation in crystalline solids. However, despite Lorentz invariance, space and time are not completely interchangeable, as evidenced by their different signs in the metric tensor. Moreover, by its very own definition, the ground state or any equilibrium state of a closed quantum system does not vary with time and  Wilczek's original idea was shown to be unfeasible\,\cite{NoTimeCrystal1,NoTimeCrystal2,NoTimeCrystal3,OldTimeCrystal3,NoTimeCrystal5}. Nevertheless the idea of time crystals as new phase of matter, has generated much interest over the past decade.
More recent studies have established that time crystals can emerge under proper conditions. It is now widely accepted that time crystals can 
be realized in out-of-equilibrium systems\,\cite{OldTimeCrystal1,OldTimeCrystal2,OldTimeCrystal3,NewTimeCrystal1,NewTimeCrystal2,NewTimeCrystal3} and particularly in the presence of a periodic driving field. 
%Consensus has also grown on a set of criteria that need to be satisfied by a state to be classified as a time crystal\,\cite{RefNote}, broadening the scope of this novel state of matter from its original definition.

Time crystals have been theoretically studied and experimentally reported in a range of systems, including magnons\,\cite{TimeCrystal3,Magnon_Time_Crystal}, ultracold atoms\,\cite{Ultracold_Atom_Time_Crystal_Theory,TimeCrystal1}, superfluid quantum gas\,\cite{TimeCrystal1,TimeCrystal2} and qubits\,\cite{QubitTimeCrystal1,QubitTimeCrystal2,QubitTimeCrystal3}.
Different time crystals can be broadly categorized in two categories, continuous time crystal\,\cite{ContinuousTimeCrystal0,ContinuousTimeCrystal1,ContinuousTimeCrystal2,ContinuousTimeCrystal3} and discrete time crystal\,\cite{DiscreteTimeCrystal1,DiscreteTimeCrystal2,DiscreteTimeCrystal3}.
Discrete time crystalline behavior in a periodically driven system is characterized by the local properties that oscillate in time with a period which is a multiple of that of the driving field\,\cite{DiscreteTimeCrystal1,DiscreteTimeCrystal2,DiscreteTimeCrystal3,PrethermalExperiment1,PrethermalExperiment2,PrethermalExperiment3,PrethermalExperiment4,PrethermalExperiment5,FractionalTimeCrystal1,FractionalTimeCrystal2,QuasiCrystal,QuasiCrystal1,QuasiCrystal2,QuasiCrystal4,QuasiCrystal5,QuasiCrystal6,DrivenDissipative1,DrivenDissipative2,DrivenDissipative3,DrivenDissipative4,DrivenDissipative5,Ultracold_Atom_Time_Crystal_Theory,TimeCrystal1,Magnon_Time_Crystal,TimeCrystal3,ArchimedeanScrew,TimeCrystal1,TimeCrystal2,TimeCrystal3}. 
%In addition to spontaneous breaking of time translation symmetry, consensus has also grown on a set of criteria that need to be satisfied by a state to be classified as a time crystal\,\cite{RefNote}.
%The criterion to observe oscillation over long time as opposed to infinite time, arises from the fact that, 
In many cases, the driving field injects energy into the system that eventually leads to thermalization. The periodic behaviours before thermalization are known as pre-thermal time crystals\,\cite{PrethermalExperiment1,PrethermalExperiment2,PrethermalExperiment3,PrethermalExperiment4,PrethermalExperiment5}.
Conversely, if the driving frequency is much larger than the local energy scales or if heat generated during thermalization can be dissipated, a driven dissipative time crystal can form because thermalization takes a long time\,\cite{DrivenDissipative0,DrivenDissipative1,DrivenDissipative2,DrivenDissipative3,DrivenDissipative4,DrivenDissipative5}.
%class of systems shown to exhibit time crystalline behavior include 
For Floquet many body localized (MBL) systems\,\cite{DiscreteTimeCrystal1,DiscreteTimeCrystal2,DiscreteTimeCrystal3}, where absence of coupling between different energy eigenstates prevent thermalization of the states, a more robust long-lived time crystal can be realized. However, MBL phase requires a strong disorder, which is practically not feasible to prepare experimentally.
Thus, a research effort is still ongoing to find new more efficient and effective ways for stabilizing the manybody phase time crystal which is much more unstable than its' counterpart space crystals.

\begin{figure}[tb]
\includegraphics[width=0.3\textwidth]{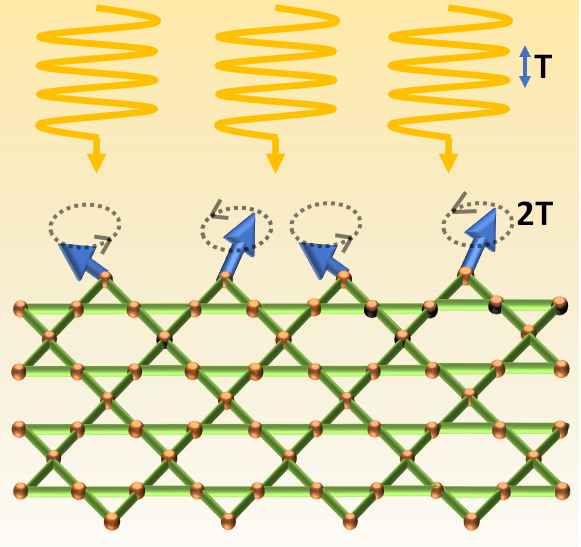} % this command will be ignored
\caption{The realization of a discrete time crystal made of topological edge magnons. At the edge spins oscillate with a period twice that of external EM field, breaking discrete time translational symmetry. The phases of oscillation for neighbouring sites are opposite as the magnons are amplified at $k=\pi$ point.}
\label{fig::Schematic}
\end{figure}

In this work, 
%for the first time 
we show that a discrete time crystal can emerge in the topological $\pi$-Berry phase protected magnon edge state of a quantum magnet driven by a periodic field in absence of any time reversal symmetry breaking interactions\,(see Fig.\,\ref{fig::Schematic}). 
%Dzyaloshinskii-Moriya interaction\,(DMI).
The topological protection of the edge state strongly reduces the scattering of edge magnons. %which in turn prevents thermalization of the magnons.
%This in turn prevents thermalization of magnons, and facilitates the stabilization of time crystalline behavior.
%Although the system eventually thermalizes, this topological protection facilitates the stabilization of coherent magnons in the pre-thermal regime. 
%absence of true many body localization results in the eventual thermalization of the magnons, the prethermal regime of the coherent magnon stabilizes the time crystal behavior, enabling us to rid the verison protected by large disorder with strong interaction. 
In contrast to Floquet MBL, our proposal avoids the need for strong disorder, which stands in the way of experimental realization in larger systems\,\cite{NewTimeCrystal3}.We work with a realistic microscopic Hamiltonian that captures the low-energy magnetic properties of the quasi-2D quantum magnets haydeeite and Cu(1,3-bdc).
 The topological time-crystal considered in this study should not be confused with the Floquet topological time crystal studied in  references Ref.\,\cite{TopologicaTimecrystal1, TopologicalTimecrystal2}, because the topology in the latter systems is out of equilibrium phenomena, whereas the topology in our system is intrinsic to the system which ensures the stability of time crystal.
While we have demonstrated the emergence of time crystalline state in magnons on the kagome lattice, the results are applicable to other bosonic quasiparticles like
%Despite of very specific results of time crystal in magnonic system in kagome lattice, the results are still important and applicable in wide range of condensed matter systems like 
phonons, plasmons, polaritons etc. where parametric amplification has been demonstrated %due to applicability of parametric amplification\%
,\cite{POLARITON_PARAMETRICAMPLIFICATION_1,POLARITON_PARAMETRICAMPLIFICATION_2,POLARITON_PARAMETRICAMPLIFICATION_3,POLARITON_PARAMETRICAMPLIFICATION_4, PHONON__PARAMETRICAMPLIFICATION_1, PHONON__PARAMETRICAMPLIFICATION_2, PHONON__PARAMETRICAMPLIFICATION_3, PHONON__PARAMETRICAMPLIFICATION_4,PLASMON_PARAMETRICAMPLIFICATION_1,PLASMON_PARAMETRICAMPLIFICATION_2,PLASMON_PARAMETRICAMPLIFICATION_3} and topological edge states are also present\,\cite{POLARITON_TOPOLOGICALEDGE_1,POLARITON_TOPOLOGICALEDGE_2,POLARITON_TOPOLOGICALEDGE_3,POLARITON_TOPOLOGICALEDGE_4,POLARITON_TOPOLOGICALEDGE_5,POLARITON_TOPOLOGICALEDGE_6,PHONON_TOPOLOGICALEDGE_1,PHONON_TOPOLOGICALEDGE_2,PHONON_TOPOLOGICALEDGE_3,PHONON_TOPOLOGICALEDGE_4,PLASMON_TOPOLOGICALEDGE_1,PLASMON_TOPOLOGICALEDGE_2,PLASMON_TOPOLOGICALEDGE_3,PLASMON_TOPOLOGICALEDGE_4,PLASMON_TOPOLOGICALEDGE_5}.

%The proposed emergent magnon time crystal can be understood as a pre-thermal time crystal from a driven-dissipative system that is further stabilised by the topological structure of the magnon band.
%the time crystal can be stabilized via dissipation and thus the emergent time crystal belongs to either pre-thermal or driven dissipative category.

%\section{\label{sec2}Results}

\begin{figure}[tb]
\includegraphics[width=0.5\textwidth]{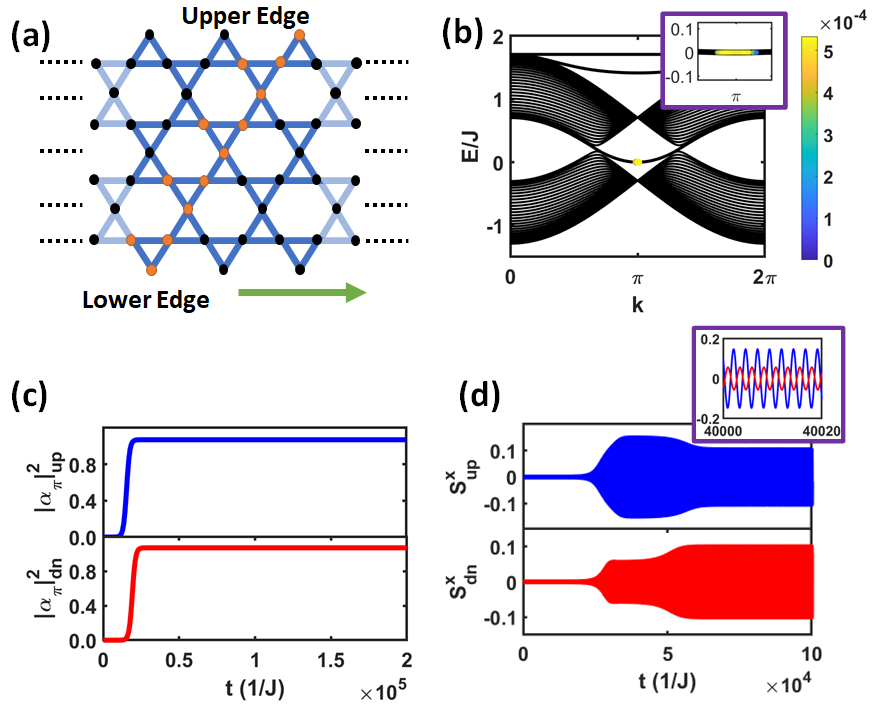} % this command will be ignored
\caption{(a) Schematic of a kagome ferromagnetic system with a finite number of sites along width (orange sites) and the system is periodic along the green arrow after certain number of sites. (b) The magnon band structure is shown in black. The yellow dots denote the amplified topological edge magnonic states. Inset shows the magnified picture of the band near $k=\pi$ and the color code describes the value of positive imaginary part of eigenvalues. (c) The number of magnons at $k=\pi$ as a function of time for upper (blue) and lower edge (red) states, indicating parametric amplification of edge magnons. (d) Time crystal made of topological edge magnons reflected in oscillation of in-plane spin component $S^x$ at a site at upper\,(blue plot) and lower edges\,(red plot). Inset shows the magnified figure within a particular time limit. The period of oscillation is twice that of the external EM field. The parameters used for all the plots are $J=1.0$, $\gamma=5\times 10^{-4}$, $\eta=9\times 10^{-4}$, $\Omega=5.1716$, $p_0=1.0$, $E_0^x=0.0$, $E_0^y=0.002$.}
\label{fig::TimeCrystal}
\end{figure}

%\subsection{\label{sec2a}Discrete time-crystal of edge magnons}
\section{\label{Section::II}Discrete time crystal} %We consider the ferromagnetic Heisenberg model on a kagome lattice in the presence of the external magnetic field:
%\begin{equation}
%    \pazocal{H}_0=-J\sum_{ij} \hat{S}_i\cdot\hat{S}_j-B_z\sum_{i} \hat{S}_i^z,
%    \label{eq::SpinHamiltonian}
%\end{equation}
%where $J$ is the Heisenberg interaction strength, $B_z$ is the magnetic field and $\hat{S}_i$ is the spin operator at the $i$-th site. 
%Eq.(\ref{eq::SpinHamiltonian}) resembles the Hamiltonian of kagome magnets such as Haydeeite\,\cite{Material1} and  Cu(1,3-bdc)\,\cite{Material2}. 
We consider the ferromagnetic Heisenberg model on a kagome lattice,
\begin{equation}
\pazocal{H}_0=-J\sum_{ij} \hat{S}_i\cdot\hat{S}_j. 
\label{Eq::UnperturbedHamiltonian}
\end{equation}
The low energy magnon excitations above the ferromagnetic ground states are described by the linear spin-wave theory: $\hat{S}_i^{+}=\sqrt{2S}\hat{a}_i$, $\hat{S}_i^{-}=\sqrt{2S}\hat{a}^{\dagger}_i$, $\hat{S}_i^z=S-\hat{a}_i^\dagger \hat{a}_i$,
where $S$ denotes the magnitude of the spin, and $\hat{a}_i^\dagger (\hat{a}_i)$ creates (annihilates) a magnon at site $i$. Application of the above transformation to $\pazocal{H}_0$ %Eq.~(\ref{eq::SpinHamiltonian}) 
yields a tight binding magnon Hamiltonian where the interactions are neglected. The resulting band structure for a ribbon geometry (Fig.\ref{fig::TimeCrystal}(a)) is shown in Fig.\,\ref{fig::TimeCrystal}(b).
The bulk bands carry a non-trivial quantized $\mathbb{Z}_2$ topological invariant (Zak-phase or $\pi$-Berry phase), and contain nearly flat topological edge states between the projected Dirac points\,\cite{BerryPhase1,BerryPhase2}.
Any effective time reversal symmetry breaking terms in the Hamiltonian, such as the Dzyaloshinskii-Moriya interaction(DMI), would open up a gap in the magnon spectrum at the Dirac points\,\cite{EffectiveTimeReversalSymmetry}
and imparts dispersion to the edge states at $k=\pi$, destroying discrete time crystalline behavior that is discussed later in Section.\,\ref{Section::III}.
%Presence of DMI would result in a cha
%Moreover, trimarization of kagome lattice would result in a \,\cite{Trimarization}.

As bosons not subject to the Pauli exclusion principle, magnons normally populate the bottom of the band, far from the edge states. However, recent studies have shown that edge state magnons can be controllably amplified at arbitrary energies by tailored EM waves\,\cite{Amplification}.
%In particular, the edge state can be populated by magnons using this approach\,\cite{Amplification}.
The EM field with amplitude $\boldsymbol{E}$ and frequency $\Omega$ couples to the magnetic insulators 
via polarization\,\cite{PolarizationOperator1,PolarizationOperator2,PolarizationOperator3} as %and the coupling Hamiltonian is given by,
\begin{align}
    H_c &=\cos(\Omega t)\boldsymbol{E}\cdot\sum_{\left\langle i,j\right\rangle} \boldsymbol{P}_{ij}
    \label{Eq::Coupling}
\end{align}
 where $\boldsymbol{P}_{ij}$ is the polarization operator of magnetic insulator. The relevant terms in $\boldsymbol{P}_{ij}$ that contribute to magnon amplification consists of bilinear spin operators on the nearest neighbor bonds\,(see Appendix.\,\ref{Appendix::A}),
 %the polarization and the most significant terms are bilinear in spin operators with spins on the nearest-neighbour bonds.
 %The bilinear spin terms which contributes in amplification is given in the supplementary of the reference Ref.\,\citep{Amplification},
 \begin{equation}
     \mathbf{P}_{ij}\approx \boldsymbol{p}_{0,ij} \left(\mathbf{S}_i\cdot\mathbf{Q}_{ij}\right)
    \left(\mathbf{S}_j\cdot\mathbf{Q}_{ij}\right).
    \label{eq::Polarization}
 \end{equation}
 %Appendix\,\ref{appendixA}.
 Other polarization terms are not important for this study 
 %because those terms will be 
 as they will be neglected in the rotating wave approximation\,\cite{Amplification}.
 Furthermore, we have neglected the effect of the magnetic field component of electromagnetic wave for the following reasons. The magnetic field couples with the system in a form of Zeeman coupling which carries a term proportional to $\cos\left(\Omega t\right)$ and only able to couple with the magnons of energy $\Omega$. Thus the edge magnons remain unaffected which has a frequency of $\Omega/2$ (see later in this section). Bulk magnons are also less affected by magnetic field, as discussed in Appendix.\,\ref{Appendix::F}.
 
The equation of motion for the magnon field $\tilde{\alpha}_k$ is given by\,(see Supplementary Section.\,B for more details),
{\small
\begin{equation}
    \frac{d}{dt}
    \begin{pmatrix}
    \tilde{\alpha}_k^*\\ 
    \tilde{\alpha}_{-k}
    \end{pmatrix}
    =
    \mathrm{i}
    \begin{pmatrix}
    \tilde{\epsilon}_k-\mathrm{i}\frac{\gamma\mathbb{I}+\eta\left|\alpha_k\right|^2}{2} & 
    \frac{\left[\tilde{H}_c\right]_{12}}{2} \\
    -\frac{\left[\tilde{H}_c\right]_{21}}{2} &
    -\tilde{\epsilon}_{-k}-\mathrm{i}\frac{\gamma\mathbb{I}+\eta\left|\alpha_k\right|^2}{2}
    \end{pmatrix}
    \begin{pmatrix}
    \tilde{\alpha}_k^*\\ 
    \tilde{\alpha}_{-k}
    \end{pmatrix},
    \label{eq::EOM}
\end{equation}
}
where $(\tilde{\alpha}^*_k \,\,\tilde{\alpha}_{-k})$ represent the 
%$\tilde{\alpha}_k$ is a column matrix with elements as 
magnon fields $\left\langle\hat{\tilde{a}}_{n,k}\right\rangle$; 
%and $\hat{\tilde{a}}_{n,k}$ is the magnon annihilation operator of $n$-th band at $k$-point; 
$\tilde{\epsilon}_{k}$ is a diagonal matrix with elements $\epsilon_{n,k}-\frac{\Omega}{2}$, where $\epsilon_{n,k}$ is the energy eigenvalue;
%is the diagonal matrix of eigenvalues of matrix $H_0(k)$ subtracted by $\frac{\Omega}{2}$; 
$\gamma$ and $\eta$ are phenomenological linear and non-linear damping constants; 
$\mathbb{I}$ is the identity matrix and $\left|\alpha_k\right|^2$ is the diagonal matrix with entries $\left|\left\langle\hat{\tilde{a}}_{n,k}\right\rangle\right|^2$.%{\small $\left(\left|\left\langle\hat{\tilde{a}}_{1,k}\right\rangle\right|^2,\left|\left\langle\hat{\tilde{a}}_{2,k}\right\rangle\right|^2,\,...,\,\left|\left\langle\hat{\tilde{a}}_{N,k}\right\rangle\right|^2\right)$}. 
The subscripts $n$ and $k$ are band-index and reciprocal space point respectively.
Moreover, $\left[\tilde{H}_c\right]_{12}$ is the off-diagonal elements of the coupling matrix in eigenbasis\,(see Appendix.\,\ref{Appendix::B}).
The square matrix on the right hand side of Eq.\,\ref{eq::EOM} is the dynamical matrix with complex eigenvalues (for $\eta=0$).
The real and imaginary parts of the eigenvalues represent the energy and lifetime of the magnon respectively.
In absence of EM coupling ({\footnotesize $\left[\tilde{H}_c\right]_{12}\approx O_{N\times N}$}), the imaginary part of eigenvalues is negative indicating magnon decay.
However as the amplitude of EM field increases the imaginary part of some of the eigenvalues satisfying $\epsilon_{n,k}+\epsilon_{n,-k}\approx \Omega$ become positive. This indicates the onset of spontaneous amplification of magnons.
The yellow dots in the band structure (Fig.\,\ref{fig::TimeCrystal}(b)) are the eigenvalues with positive imaginary part.

The solution of Eq.\ref{eq::EOM} describes amplified coherent magnons above a cutoff amplitude of EM field\,\cite{CoherentState1,CoherentState2,TimeCrystal1,TimeCrystal2}.
Fig.\,\ref{fig::TimeCrystal}(c) shows the amplified coherent magnons population for the edge states of upper and lower edges at $k=\pi$ as a function of time.
The presence of the non-linear damping suppresses the exponential increase of the magnon number and the system reaches a steady state.

While the number of magnons 
({\small $\left|\left\langle\hat{\tilde{a}}_{n,k}\right\rangle\right|^2$}) are identical in the rotating and the lab frames in the steady state (see Fig.\,\ref{fig::TimeCrystal}), the field $\left\langle\hat{\tilde{a}}_{n,k}\right\rangle$ oscillates in time.
%\begin{equation}
%    \left\langle \hat{\tilde{a}}_{i,k}(t)\right\rangle_{\text{lab}}
%    =
%    \left\langle\hat{\tilde{a}}_{i,k}(t)\right\rangle_{\text{rot}}
%    \exp(\mathrm{i}\frac{\Omega}{2} t),
%\end{equation}
Specifically when the pair of amplified magnons satisfy $\epsilon_{n,k}=\epsilon_{n,-k}=\Omega/2$, the steady state expectation values for the field in the rotating frame is independent of time i.e. $\left\langle\hat{\tilde{a}}_{n,k}(t)\right\rangle_{\text{rot}}^s=\left\langle\hat{\tilde{a}}_{n,k}\right\rangle_{\text{rot}}^s$\,(see Appendix.\,\ref{Appendix::D}). Thus the fields in the two frames are related as,
\begin{equation}
    \left\langle
    \hat{\tilde{a}}_{n,k}(t)\right\rangle_{\text{lab}}^{\text{s}}
    \approx
    \left\langle\hat{\tilde{a}}_{n,k}\right\rangle_{\text{rot}}^{\text{s}}
    \exp(\mathrm{i}\frac{\Omega}{2} t),
    \label{eq::Oscillation2}
\end{equation}
where the superscript ``s" denotes steady state expectation value.
%
%Moreover it can be shown if a simple two energy levels with energies $\omega_k$ and $\omega_{-k}$ are coupled according to Eq.\,\ref{eq::EOM} then the time dependency of the field $\left\langle\hat{\tilde{a}}_{i,k}\right\rangle$ in rotating frame in steady state is given by,
%\begin{equation}
%    \left\langle\hat{\tilde{a}}_{i,k}(t)\right\rangle_{\text{rot}}^\text{s}
%    =
%    \left\langle\hat{\tilde{a}}_{i,k}\right\rangle_{\text{rot}}^\text{s}
%   \exp(\mathrm{i}\frac{\omega_k-\omega_{-k}}{2} t),
%    \label{eq::Oscillation1}
%\end{equation}
%where superscript s denotes steady state.
%Thus for amplification of magnons with nearly same energies would result in a oscillation with half of the frequency of light in steady state,

The equation of motion Eq.\,\ref{eq::EOM} has a $\mathbb{Z}_2$ symmetry  $\hat{\tilde{a}}_{n,k}\rightarrow -\hat{\tilde{a}}_{n,k}$.
Above a critical amplitude of the EM field, the amplified magnon field at the edges the system spontaneously breaks the $\mathbb{Z}_2$ symmetry by acquiring a finite, non-zero {\small $\left\langle\hat{\tilde{a}}_{n,k}\right\rangle_{\text{lab}}^{\text{s}}$} that oscillates in time with a period which is twice that of the driving EM field.
Thus a discrete time crystal of edge state magnons is formed via amplification\,\cite{TimeCrystal1,TimeCrystal2} that breaks the discrete time translational symmetry spontaneously.

This time crystalline behavior can be experimentally observed by measuring the transverse magnetization at the edges, i.e., the spin components $\hat{S}^x_i$ and $\hat{S}^y_i$ -- the spin component $\hat{S}^z_i$ is constant, because it is related to the number of magnons {\small $\left|\left\langle\hat{\tilde{a}}_{i,k}\right\rangle\right|^2$}, which is invariant in time in the steady state. 
The x-component of the spin, $\left\langle \hat{S}^x_i\right\rangle$,  is given in terms of the fields $\left\langle \hat{\tilde{a}}_{n,k}\right\rangle_{\text{rot}}^{\text{s}}$ as
\begin{widetext}

\begin{equation}
    \left\langle\hat{S}_i^x\right\rangle = \sqrt{\frac{S}{2N_x}}
    \left[
    \sum_{k>0,n} \left[U_1^\dagger\right]_{in} \left\langle \hat{\tilde{a}}_{n,k}\right\rangle_{\text{rot}}^{\text{s}} e^{\frac{i\Omega t}{2}} e^{-ikx_i}
    +
    \sum_{k>0,n} \left[U_2^\dagger\right]_{in} \left\langle \hat{\tilde{a}}_{n,-k}\right\rangle_{\text{rot}}^{\text{s}} e^{\frac{i\Omega t}{2}} e^{ikx_i}
    +
     \left[U_1^\dagger\right]_{in} \left\langle \hat{\tilde{a}}_{n,0}\right\rangle_{\text{rot}}^{\text{s}} e^{\frac{i\Omega t}{2}}
     +
     \text{\small{H.C.}}
    \right]
    \label{eq::Sx}
\end{equation}

\end{widetext}

Fig.\,\ref{fig::TimeCrystal}(d) demonstrate the oscillation of $S^x_i$ at a site at the upper edge (blue) and the lower edge  (red). The different $k$-points arrive at steady state at a different time\,(see Appendix.\,\ref{Appendix::C}) and so $S^x_i$ in Fig.\,\ref{fig::TimeCrystal}(d) modulates transiently and reaches a steady state when all the amplified $k$-points do.
%It is noticeable that the time taken to reach steady state is different for different $k$-points (compare Fig.\,\ref{fig::TimeCrystal}(d) and (e)), thus the $S^x_i$ contributed by all the $k$-points exhibits a transient amplitude modulation (see Fig.\,\ref{fig::TimeCrystal}(f)) and reaches a steady state when all the amplified $k$-points come to the steady state.
 The oscillation amplitudes of $\left\langle \hat{S}_i^x\right\rangle$ at both edges are nearly identical, but they are not exactly equal, since it is a superposition of several fields $\left\langle \hat{\tilde{a}}_{n,k}\right\rangle_{\text{rot}}^{\text{s}}$\,(see Eq.\,\ref{eq::Sx}); while the magnitude of the fields $\left\langle \hat{\tilde{a}}_{n,k}\right\rangle_{\text{rot}}^{\text{s}}$ at two edges are the same, the phases are not.
Moreover, the amplitude of oscillation varies with different simulations due to the random starting conditions representing the vacuum fluctuations\,\cite{CoherentState1,CoherentState2}.
Finally, time crystalline behavior also holds for the long-range order in spatial directions due to the coherence of the pumped magnon at $k=\pi$. 
Since the amplification of magnons extends over a finite momentum range around $k=\pi$, a spatial modulation in the amplitude of oscillation is expected.

%\begin{widetext}

%\end{widetext}

%\subsection{\label{sec2a}Stability of time crystal}

\section{\label{Section::III}Symmetry Protection}

\begin{figure}[t]
\includegraphics[width=0.5\textwidth]{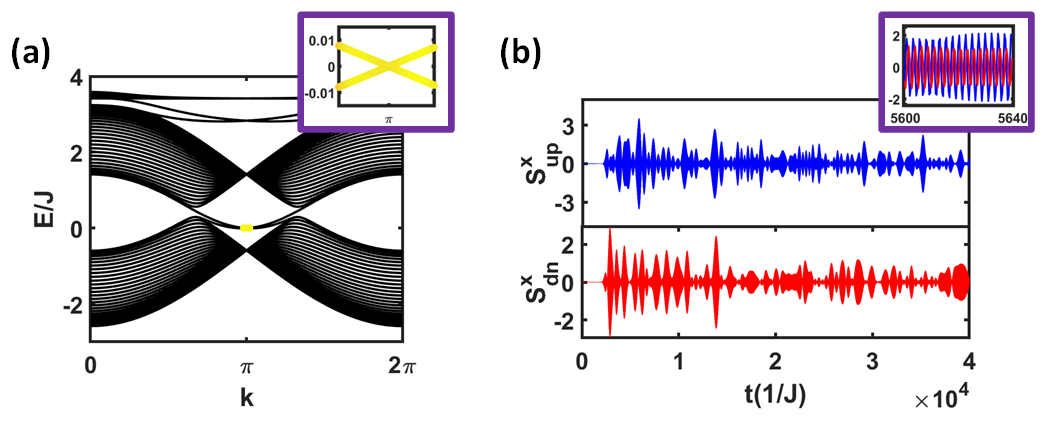} % this command will be ignored
\caption{(a) The magnon band structure in presence of effective time reversal symmetry breaking DMI. The yellow dots denote amplified magnon eigenstates. Inset shows the magnified band near $k=\pi$. (b) A chaotic oscillation of spin component $S^x$ at a site at the upper (blue plot) and lower edges (red plot) are shown, indicating the destruction of time crystal. Insets show the magnified figures of the corresponding figures within a particular time limit. The parameters used for all the plots are $J=1.0$, $D=0.05$,  $B_z\rightarrow 0^+$, $\gamma=5\times 10^{-4}$, $\eta=9\times 10^{-4}$, $\Omega=5.1716$, $p_0=1.0$, $E_0^x=0.0$, $E_0^y=0.02$.}
\label{fig::TimeCrystal2}
\end{figure}
Although Heisenberg ferromagnet breaks time-reversal symmetry $\tau$, the system still preserves effective time reversal symmetry $\tau_{\text{eff}}=\pazocal{R}_\pi(\boldsymbol{n})\tau$, where $\pazocal{R}_\pi(\boldsymbol{n})$ is $\pi$-rotation of spins around $\boldsymbol{n}$-axis. This effective symmetry is useful to describe the system in the language of magnons\,\cite{EffectiveTimeReversalSymmetry}.
 The breaking of effective time reversal symmetry, which depends on both the direction of DMI and spin-moments of ferromagnetic ground state, is ubiquitous in many real quantum magnets, resulting in dispersive edge states.
As a result, 
%the pair of amplified points will not have the same energy, 
the condition, $\epsilon_{n,k} = \epsilon_{n,-k}$ for a pair of amplified magnons, as assumed above, is broken.
%as in Ref.\,\cite{Amplification}.
Then, $\left\langle \hat{\tilde{a}}_{n,k}\right\rangle_{\text{rot}}^s$ becomes time dependent and according to Eq.\,\ref{eq::Oscillation2}, the period of oscillation of the fields at a particular $k$-point will no longer be exactly twice the period of the external EM field. 
For a finite system, there will be a finite number of amplified points around $k=\pi$; adding over the fields at a few amplified points according to Eq.\,\ref{eq::Sx} would result in an oscillation of $S^x$ that is incommensurate with the external field, resulting in a quasi-time crystal\,\cite{QuasiCrystal}.
%For a finite size system there would be a finite number of $k$-points nearby $k=\pi$ that would be amplified, thus summing over the fields of few amplified $k$-points according to Eq.\,\ref{eq::Sx} would result in a oscillation of $S_x$ which is incommensurate with the external field, resulting in a quasi-time crystal.
%Summing over the fields over few $k$-points according to Eq.\,\ref{eq::Sx} would result in a quasi-time crystal for a finite size system.
In the thermodynamic limit, when the number of $k$-points in the vicinity of $k=\pi$ diverges, the oscillation would become chaotic in nature, destroying the time crystal-like behaviour.

Here we demonstrate the absence of time crystal, in presence of time-reversal symmetry breaking DMI.
We would like to emphasize that not all DMI breaks time-reversal symmetry. Breaking of time reversal symmetry depends both on the direction of DMI and direction of spin moments of a ferromagnetic ground state\,\cite{EffectiveTimeReversalSymmetry}.
%For further details, please refer to the "Symmetry analysis" section of reference Ref.\,\cite{EffectiveTimeReversalSymmetry}.
The model Hamiltonian in absence of time reversal symmetry becomes,
\begin{equation}
    \pazocal{H}=-J\sum_{ij} \hat{S}_i\cdot\hat{S}_j+D\sum_{ij}\hat{z}\cdot (\hat{S}_i\times\hat{S}_j)-B_z\sum_i \hat{S}_i^z,
\end{equation}
where $D$ and $B_z$ are DMI and magnetic field perpendicular to 2D-lattice plane.
The breaking of time-reversal symmetry opens up a gap in the bulk-magnon band and we get dispersive edge magnon state.
Thus the amplification is not only limited to the $k=\pi$ but other momentum points such that the relationship $\omega_k+\omega_{-k}=\Omega$ is satisfied, which is shown in Fig.\,\ref{fig::TimeCrystal2}(a).
Moreover the spin oscillation at the edges become chaotic in nature as shown in Fig.\,\ref{fig::TimeCrystal2}(b).
The reason behind this chaotic oscillation can be understood from the following simple model with only two coupled modes with energies $\omega_k$ and $\omega_{-k}$,
\begin{equation}
    \pazocal{H}=\omega_k \hat{a}_k^\dagger \hat{a}_k + \omega_{-k} \hat{a}_{-k}^\dagger \hat{a}_{-k} +2\epsilon \cos(\Omega t) \left(\hat{a}_k^\dagger \hat{a}_{-k}^\dagger + \hat{a}_k \hat{a}_{-k}\right).
\end{equation}
It can be shown that the fields in lab frame have the following time dependence (see Appendix.\,\ref{Appendix::D}),
\begin{equation}
    \alpha_{\pm k}^{\text{lab}}=A_{\pm}(t) e^{\mp i\frac{(\omega_k-\omega_{-k})t}{2}}
    e^{-i\frac{\Omega}{2}t},
\end{equation}
where $A_\pm(t)$ is amplitude modulation due to amplification which becomes time independent in the steady state.
From this equation, we can conclude the oscillation frequency due to parametric amplification is half of the driving field only when the coupled state which is amplified have equal energies $\omega_k=\omega_{-k}$, otherwise the frequency of oscillation is different.
Moreover, when effects of oscillation from many coupled oscillators with $\omega_k\neq \omega_{-k}$ are superimposed as in equation Eq.\,\ref{eq::Sx}, then the resultant oscillation will be chaotic as in Fig.\,\ref{fig::TimeCrystal2}(b).

\section{\label{Section::IV}Stability of the time crystal}

\begin{figure}[tb]
\includegraphics[width=0.5\textwidth]{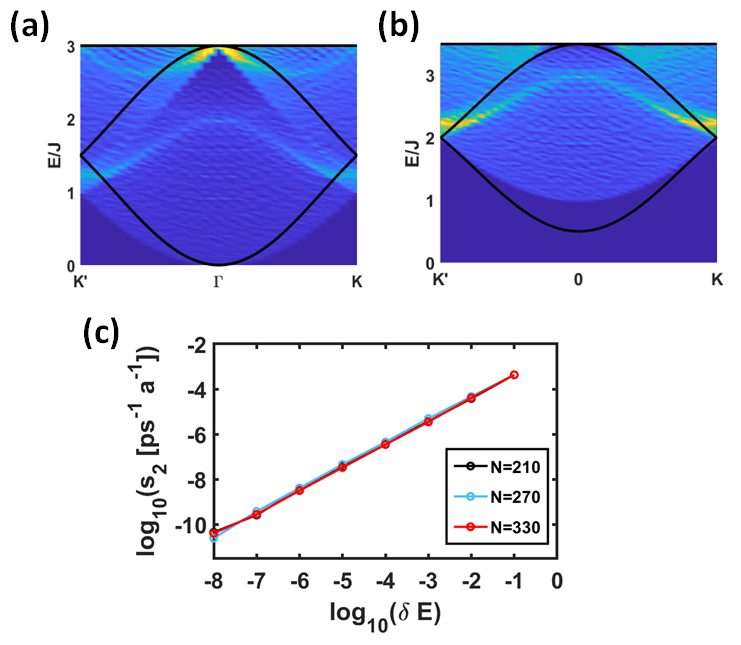} % this command will be ignored
\caption{ The two magnon density of states\,(bluish-yellow color map) and band structure\,(black line) for (a) zero magnetic field and (b) finite magnetic field $B_z=JS$. With increasing magnetic field, the overlap between two magnon continuum with the band structure decreases, signifying fewer magnon non-conserving scattering. (c) The two-magnon scattering rate as a function of band broadening for different system widths. The scattering rate is $s_2\leq 10ns^{-1}a^{-1}$ for an ideal band broadening $\delta E\leq 10^{-2}JS$, indicating a low scattering rate compared with ideal magnon lifetime $\tau_{\text{life}}\leq 1ns$. Thus secondary amplification of bulk magnon is negligible.} %with fixed length $N_x=100000$.}
\label{fig::Stability}
\end{figure}

Discrete time crystals of the amplified edge magnons are stable unless bulk eigenmodes with a significant overlap with the edge modes is amplified.
% due to interaction-induced magnon scattering.
The choice of the kagome ferromagnet is preferable because the energy spectrum around edge state is asymmetrical\,(see Fig.\,\ref{fig::TimeCrystal}(b)), resulting in suppression of bulk magnon amplification\,(see Appendix.\,\ref{Appendix::E}, where a comparison with honeycomb ferromagnet is provided).
%, if the $y$-polarized light is applied\,
%(see supplemental material of Ref.\,\cite{Amplification}).
However, edge magnon scattering can excite other bulk magnon eigenmodes and contribute to the oscillation of spin according to Eq.\,\ref{eq::Sx}, which may become the reason of instability of time crystal.

%represented by the Hamiltonian terms higher than the quadratic order, which can be obtained via Holstein-Primakoff transformation of spin Hamiltonian Eq.\,\ref{eq::SpinHamiltonian} and $1/S$-expansion.
%The scattering between magnons can be divided into two categories: number non-conserving and conserving scattering.

The Hamiltonian $\pazocal{H}_0$
%(\ref{eq::SpinHamiltonian}) 
does not contain any magnon non-conserving terms, but such terms may arise in the presence of spin anisotropy in many real quantum magnets.
We have calculated the bulk band structure and momentum resolved two-magnon density of states, 
\begin{equation}
    D_\bk(\omega)=\frac{1}{N} \sum_{n,n',\boldsymbol{\Delta}}
    \delta(\omega-\omega_{n,\boldsymbol{\Delta}}-\omega_{n',\bk-\boldsymbol{\Delta}})
\end{equation}
for a system with toroidal boundary condition\,(Fig.\,\ref{fig::Stability}(a), (b)). 
When the energy of an eigenstate matches that of the two magnon continuum, it can decay into two magnons with lower energies.
%{\color{blue}The two magnon continuum energy scales as twice the negative of Zeeman term ($2B_zS$) induced via the magnetic field $B_z$ applied in the direction of ferromagnetic spin, while that of the magnon bands scale as $B_zS$.}
The two magnon continuum energy scales as twice the negative of Zeeman term ($2B_zS$) of a longitudinal magnetic field $B_z$, while that of the magnon bands scale as $B_zS$.
Hence for magnetic fields $B_z S>E_{\text{edge}}$ the edge states are energetically separated from the 2-magnon continuum and cannot decay via this channel.
%there will be no overlap between two magnon continuum and edge states, indicating that the edge states cannot decay in this channel.
%into two lower energy magnons.
At even higher fields, $B_z>6J$ the two-magnon continuum gets separated in energy from the magnon band structure, implying that two edge magnons cannot combine and produce a higher energy magnon. 
As the magnetic field increases, the higher order magnon non-conserving processes will disappear faster than the two magnon decay processes.
Thus an external magnetic field can suppress the magnon non-conserving scattering in the system.

Magnon conserving scattering processes can not be eliminated by an external field, and are always present in any spin Hamiltonian.
We calculated the scattering rate of magnons due to two-magnon scattering\,(quartic terms of magnon Hamiltonian) using Fermi golden rule,
\begin{equation}
    s_2=\frac{2\pi}{\hbar} \sum_f\sum_i\left|\squeezeB{f}{\pazocal{H}_{0,\text{int}}^{(4)}}{i}\right|^2 \delta(E_f-E_i),
    \label{eq::FermiGoldenRule}
\end{equation}
where $|f\rangle$ and $E_f$ ($|i\rangle$ and $E_i$) denote the state and energy of final (initial) state respectively. $\pazocal{H}_{0,\text{int}}^{(4)}$ is the quartic term of the magnon Hamiltonian which is obtained by Tailor series expansion of Holstein-Primakoff transformation,
\begin{align}
        \pazocal{H}_{0,\text{int}}^{(4)}&=-\frac{J\hbar^2}{4} \sum_{\left\langle ij\right\rangle}
    \left[ 4\hat{a}_i^\dagger\hat{a}_j^\dagger\hat{a}_i\hat{a}_j
    +\hat{a}_j^\dagger \hat{a}_j^\dagger \hat{a}_i\hat{a}_j\right.\nonumber\\
    &\left.+\hat{a}_i^\dagger \hat{a}_j^\dagger \hat{a}_i\hat{a}_i
    +\hat{a}_i^\dagger\hat{a}_j^\dagger \hat{a}_j\hat{a}_j+\hat{a}_i^\dagger\hat{a}_i^\dagger\hat{a}_i\hat{a}_j
    \right],
\end{align}
where higher order terms of the expansion are neglected which have amplitudes of order $\pazocal{O}\left(\frac{1}{S^n}\right)$, $n\in\mathbb{I}\geq 1$.
%Scattering processes follow not only energy conservation explicitly visible in the delta function $\delta(E_f-E_i)$, but also momentum conservation implicit in the interaction Hamiltonian $\pazocal{H}_{0,\text{int}}^{(4)}$.
%The scattering processes not only follow the energy conservation explicitly noticeable due the delta function in Eq.\,\ref{eq::FermiGoldenRule}, but also follows the momentum conservation implicitly embedded into the interaction Hamiltonian $\pazocal{H}_{0,\text{int}}^4$.
For simplicity, we have restricted our calculation to the upper-edge states and considered only the scattering from the points $k=\pm \pi$ and $k=\pm\pi\pm\delta k$ (where $\delta k=0.0628$).
%$k=\pi$ and $k=-\pi$, $k=\pi$ and $k=\pi+\delta k$, and $k=\pi$ and $k=\pi-\delta k$ points.
To numerically calculate the scattering rate $s_2$, we have considered a finite width of energy levels $\delta E$, which physically implies band broadening.
The scattering rate as a function of band broadening is plotted in a logarithmic scale in Fig.\,\ref{fig::Stability}(c) for different system sizes.
It is observed that the scattering rate decreases rapidly as the band broadening decreases.
The band broadening results from magnon interactions and the only way to control the bandwidth is by controlling the density of the amplified magnons.
This is achieved by working at low temperature and low EM field intensity.
%Thus minimizing the bandwidth can reduce scattering.
%The bandwidth of magnon bands is a result of magnon interaction and higher the magnon occupation, the more intense the interaction, producing a broadened band structure.
%Therefore, band broadening can be minimized via lowering the occupation number of magnons by performing the experiment at low temperature and by amplifying small number of magnons.
Even if the scattering rate is high, as long as the life-time of magnons at the scattered state is small the behavior of the system is governed by only the selectively amplified edge state magnons\,\cite{TimeCrystal3}.
For example, the band broadening of an ideal magnon band structure $\delta E\leq 10^{-2}JS$, corresponding to scattering rate $s_2\leq 10^{-4} ps^{-1} a^{-1}=10 ns^{-1} a^{-1}$, can be interpreted as maximum one edge magnon is scattered to the bulk magnon every $10ns$, so secondary amplification of bulk magnon is not possible since their life-times are usually less than $1ns$.
%For example, the band broadening of an ideal magnon band structure $\delta E\leq 10^{-2}JS$, which corresponds to scattering rate $s_2\leq 10^{-4} ps^{-1} a^{-1}=10 ns^{-1} a^{-1}$, which can be interpreted as maximum one edge magnon is scattered to the bulk magnon every $10ns$ and because most of the time life-time of magnon is less than $1ns$, the secondary amplification of bulk magnon is not possible.
Even in the worst case scenario of large band broadening, a small density of amplified edge state magnons is preferable to create a stable time crystal. These conditions also help minimise magnon decay due to magnon-phonon scattering. Finally, edge imperfections cause broadening of the edge states, and reduce the yield of coherent magnons~\cite{Tanaka2020,Pawlak2020}.
%Hence sample quality is crucial for experimental detection \,\cite{Tanaka2020,Pawlak2020}.

%Magnons can also decay due to magnon-phonon interaction, increasing the number of phonons and generating heat in the system.
%If temperature become sufficiently high the time crystal will become destabilized due to the superposition of many magnon states with different frequencies.
%Thus a proper cryogenic temperature in the system will be required to to stabilize the time crystal.
% The edge imperfections will also cause the broadening of the edge states, and reduce the yield of coherent magnon. We therefore need well-developed synthesis methods to ensure that the experimental sample is of good quality\,\cite{Tanaka2020,Pawlak2020}.

%{\color{red}On the positive side, the edge states are topologically protected by the $\pi$-Berry phase, even in the absence of chiral symmetry.}
%{\color{blue} The low scattering rate can be argued to be an outcome of the topological protection of the edge state by the $\pi$-Berry phase, even in the absence of chiral symmetry.}
The reduced scattering is a consequence of the topological protection of the edge state.
The presence of chiral symmetry induces a flat edge mode at zero energy due to finite $\pi$-Berry phase.
Absence of chiral symmetry can result in a dispersive edge state at non-zero energy in the presence of a quantized non-zero $\pi$-Berry phase\,\cite{BerryPhase1,BerryPhase2}.
Interestingly, the $\pi$-Berry phase protected topological edge states are robust against impurities with or without chiral symmetry\,\cite{BerryPhase2}, providing additional stability to the time crystalline behavior.

%Consequently, we argue that the Berry phase protected topological edge state in kagome lattice is also robust against impurities.

%Experimentally, the edge-magnon time-crystal can be observed using multiple recently developed techniques. The Brillouin light scattering spectroscopy is useful experimental probe to detect the presence of magnon at particular frequency and wave vector\,\cite{BLS1,BLS2,BLS3,BLS4,BLS5,BLS6} and recently it has been implemented to detect the space-time crystal in the ferromagnetic insulator \ce{YIG}\,\cite{YIG_TimeCrystal}. Moreover, theoretically proposed spin Hall noise spectroscopy is a promising technique to detect the presence of edge magnons at a particular frequency\,\cite{SpinHallNoise1,SpinHallNoise2}. The direct spatial and temporal imaging of spin-wave dynamics is also possible via Kerr microscopy\,\cite{Kerr1,Kerr2}, Brillouin light scattering spectroscopy\,\cite{BLS_imaging1,BLS_imaging2,BLS_imaging3} and time resolved scanning transmission x-ray (TR-STXM) microscopy\,\cite{STXMm1,STXM0,STXM10,STXM1,STXM2,STXM3,STXM4,STXM5,STXM6,STXM8,STXM9,Magnon_Time_Crystal}. {\color{blue} Particularly the TR-STXM technique is very promising to probe spin dynamics at the edge due to its high accuracy to detect magnon dynamics with a spatial and temporal resolution of 20 nm and 50 ps respectively\,\cite{STXMm1,STXM0,STXM10,Magnon_Time_Crystal}.} Recently, this method is used to directly observe the dynamics of space-time crystal of bulk magnons in permalloy strips\,\cite{Magnon_Time_Crystal}.

\section{\label{Section::V}Experimental realisation} 
The edge-magnon time-crystals can be observed using direct spatial and temporal imaging of spin-wave dynamics via multiple recently developed techniques, such as, Kerr microscopy\,\cite{Kerr1,Kerr2}, Brillouin light scattering spectroscopy (BLS)\,\cite{BLS_imaging1,BLS_imaging2,BLS_imaging3} and time resolved scanning transmission x-ray  microscopy (TR-STXM)\,\cite{STXMm1,STXM0,STXM10,STXM1,STXM2,STXM3,STXM4,STXM5,STXM6,STXM8,STXM9,Magnon_Time_Crystal}. BLS is useful to detect magnons at a fixed frequency and wave vector\,\cite{BLS1,BLS2,BLS3,BLS4,BLS5,BLS6} and has recently been used to detect the space-time crystal in the ferromagnetic insulator \ce{YIG}\,\cite{TimeCrystal3}. Additionally, theoretically proposed spin Hall noise spectroscopy is a promising technique to detect the presence of edge magnons at a given frequency\,\cite{SpinHallNoise1,SpinHallNoise2}.
The TR-STXM, in particular, is promising for directly imaging spin dynamics at the edge due to its high accuracy in detecting magnon dynamics with a spatial and temporal resolution of 20 nm and 50 ps respectively\,\cite{STXMm1,STXM0,STXM10,Magnon_Time_Crystal}. Recently, this method has been used to observe the dynamics of space-time crystal of bulk magnons in permalloy strips\,\cite{Magnon_Time_Crystal}.

 The required estimated electric field amplitude to amplify topological edge magnons is in between $10^6$-$10^{12}$ V/m, depending on damping of edge magnons -- weaker damping requires lower intensity,(see Appendix.\,\ref{Appendix::F}).

We propose the spin-$\frac{1}{2}$ kagome ferromagnets haydeeite\,\cite{Material1} and \ce{Cu(1,3-bdc)}\,\cite{Material2_V2} as possible hosts of the discrete time crystals of edge magnons as discussed in this work.
While the Haydeeite has experimental evidence for the absence of DMI\,\cite{Material1}, the Cu(1,3-bdc) contains out-of-plane DMI that does not break any effective time reversal symmetry\,\cite{EffectiveTimeReversalSymmetry} for the ferromagnetic ground state with in-plane magnetization\,\cite{Material2}.
 The period of oscillations for the materials haydeeite and \ce{Cu(1,3-bdc)} are calculated to be $0.05$ps and $0.25$ps respectively, which are estimated using experimentally determined Heisenberg exchange interactions\,\cite{Material1,Material2_V2}.
Thus these quantum magnets are perfect candidates for realizing discrete time crystal of edge magnons.

{\it Acknowledgements} B.Y. would like to acknowledge the support from the Singapore National Research Foundation (NRF) under NRF fellowship award NRF-NRFF12-2020-0005, and a Nanyang Technological University start-up grant (NTU-SUG). D.B and P.S. acknowledge financial support from the Ministry of Education, Singapore through MOE2019-T2-2-119.

\appendix
\begin{widetext}

%\preprint{APS/123-QED}

\section{\label{Appendix::A}Polarization Operator}

The form of the polarization operator depends on the lattice symmetries and independent of the magnetic ground state.
However, for the estimation of the coefficient of the polarization operator, one requires a more fundamental electronic model.
There are various possible electronic model for the spin exchange interactions in different materials. 
One of the most simplistic electronic model is the Hubbard model.
Despite of its limitations, the simplistic Hubbard model is used to give an estimation of the coefficients of polarization operator.
The Hubbard model is given by,
\begin{equation}
      \pazocal{H}_{\text{Hubbard}}
      =
      -\sum_{ij}
      \left[
      \begin{pmatrix}
      \hat{c}_{i\uparrow}^\dagger &
      \hat{c}_{i\downarrow}^\dagger
      \end{pmatrix}
      \left(t\mathbb{I}\cos(\theta)+it\boldsymbol{n}\cdot\boldsymbol{\sigma}\sin(\theta)\right)
      \begin{pmatrix}
      \hat{c}_{j\uparrow} \\
      \hat{c}_{j\downarrow}
      \end{pmatrix}
      +
      \text{H.c.}
      \right]
      +
      U\sum_i \hat{n}_{i\uparrow}\hat{n}_{j\downarrow}
 \end{equation}
 where $t\cos(\theta)$ and $it\boldsymbol{n}\sin(\theta)$ ($\boldsymbol{n}$ is an unit vector) are the real and complex hopping amplitude of electrons on nearest neighbour bonds respectively. $U$ is the onsite Coulomb repulsion. 
 The polarization operator corresponding to the Hubbard model is\,\cite{Amplification},
\begin{equation}
     \mathbf{P}_{ij}\approx \boldsymbol{p}_{0,ij} \left(\mathbf{S}_i\cdot\mathbf{Q}_{ij}\right)
    \left(\mathbf{S}_j\cdot\mathbf{Q}_{ij}\right),
    \label{eq::Polarization}
 \end{equation}
 where the $\bold{p}_{0,ij}$ and $\boldsymbol{Q}_{ij}$ are given by,
\begin{align}
    \boldsymbol{p}_{0,ij} &=-16\theta^2 e a \frac{t^3}{U^3} (\boldsymbol{e}_{jk}-\boldsymbol{e}_{ki})
    =p_0 (\boldsymbol{e}_{jk}-\boldsymbol{e}_{ki})
    ,
    \nonumber\\
    \boldsymbol{Q}_{ij}&=\boldsymbol{n}-n^z\hat{z}
    \label{eq::OrderOfMagnitude}
\end{align}
where $e$ and $a$ are the electron charge and lattice constant respectively.
$\boldsymbol{e}_{jk}$ is a vector on nearest neighbour bonds from site-$j$ to site-$k$.
The sites $i$, $j$ and $k$ are the sites on the same triangle of the kagome lattice.
The polarization terms other than the terms in Eq.\,\ref{eq::Polarization} are not important for this study because in the diagonal basis in the rotating frame those terms are time dependent and so those terms are dropped in rotating wave approximation in Eq.\,\ref{eq::EOM}.
Thus the polarization operator is proportional to $\frac{t^3}{U^3}$.

\section{\label{Appendix::B}Derivation of equation of motion}
The Hamiltonian describing a kagome ferromagnet on a cylindrical geometry, coupled to an external EM field is given by
 %The total Hamiltonian for a system with periodic boundary condition along $x$-direction in terms of magnons in momentum space would become,
\begin{align}
    \pazocal{H}=&\frac{1}{2}\sum_{k}
    \begin{pmatrix}
    \Psi_{k}^\dagger & \Psi_{-k}
    \end{pmatrix}
    \begin{pmatrix}
    H_0(k) & O_{N\times N} \\
    O_{N\times N} & H_0(-k)^T
    \end{pmatrix}
    \begin{pmatrix}
    \Psi_{k} \\ \Psi_{-k}^\dagger
    \end{pmatrix}
    \nonumber\\
    &+
    \frac{1}{2}\cos(\Omega t)\sum_{k} 
    \begin{pmatrix}
    \Psi_{k}^\dagger & \Psi_{-k}
    \end{pmatrix}
    \begin{pmatrix}
    [H_c]_{11} & [H_c]_{12} \\
    [H_c]_{21} & [H_c]_{22}
    \end{pmatrix}
    \begin{pmatrix}
    \Psi_{k} \\ \Psi_{-k}^\dagger
    \end{pmatrix}.
\end{align}
$N$ is the number of sites along the width of the ribbon\,(see Fig.\,2(a) in main text), and $\Psi_{k}=\left(\hat{a}_{1,k},\,\hat{a}_{2,k},\,...,\,\hat{a}_{N,k}\right)^T$.% is the column matrix of magnon operators.
%respectively and $1,\dots,N$ denote the  sites along the width of the ribbon\,(see Fig.\,\ref{fig::TimeCrystal}(a)).
$O_{N\times N}$ is a null matrix.
The first and second matrices are derived from the unperturbed Hamiltonian Eq.\,\ref{Eq::UnperturbedHamiltonian} and coupling Hamiltonian Eq.\,\ref{Eq::Coupling} in the main text respectively.
The Hamiltonian $\pazocal{H}$ is first represented in the diagonal basis $\tilde{\Psi}_k=U_1(k)\Psi_k$, $\tilde{\Psi}_{-k}^\dagger=U_2(k)\Psi_{-k}^\dagger$ of matrices $H_0(k)$, $H_0(-k)^T$. 
This is followed by transforming the system from the lab-frame to rotating-frame by using the unitary operator $U(t)=\exp{\frac{i\omega t}{2}\sum_k  \tilde{\Psi}_k^\dagger \tilde{\Psi}_k}$,
\begin{equation}
    \pazocal{H}^\prime=U(t) \pazocal{H} U(t)^\dagger + i\hbar \dot{U}(t) U(t)^\dagger.
\end{equation}
Afterwards, by neglecting the time-dependent terms, we get the following effective Hamiltonian:
\begin{equation}
\pazocal{H}_{\text{eff}}=\frac{1}{2}
\sum_{k}
\begin{pmatrix}
\tilde{\Psi}_k^\dagger  & \tilde{\Psi}_{-k}
\end{pmatrix}
\begin{pmatrix}
\epsilon_k-\frac{\Omega}{2} & \frac{[\tilde{H}_{c}]_{12}}{2} \\
\frac{[\tilde{H}_{c}]_{21}}{2} & \epsilon_{-k}-\frac{\Omega}{2}
\end{pmatrix}
\begin{pmatrix}
\tilde{\Psi}_k  \\ \tilde{\Psi}_{-k}^\dagger
\end{pmatrix},
\label{eq::EffectiveHamiltonian}
\end{equation}
where $\epsilon_{k}$ and $\epsilon_{-k}$ are the diagonal matrices of eigenvalues of matrices $H_0(k)$ and $H_0(-k)^T$ respectively.
The matrix $\tilde{H}_{12}=U_1 H_{12} U_2^\dagger$ is the coupling matrix in the diagonal basis. 
%Moreover it is noticeable that the coupling matrices $H_{11}$ and $H_{22}$ representing hopping terms do not 
Only the off-diagonal terms $[\tilde{H}]_{12}$ and $[\tilde{H}]_{21}$ of the coupling Hamiltonian appear in $\pazocal{H}_{\text{eff}}$ due to the rotating wave approximation. %and that is why the only term of polarization that would result in pair-creation and annihilation terms is taken into account in Eq.\,\ref{eq::Polarization}.
The equation of motion of field $\left\langle \tilde{\Psi}_{k}\right\rangle=\tilde{\alpha}_k=\left(\left\langle\hat{\tilde{a}}_{1,k}\right\rangle,\,\left\langle\hat{\tilde{a}}_{2,k}\right\rangle,\,...,\,\left\langle\hat{\tilde{a}}_{N,k}\right\rangle\right)^T$ is given by,
{\small
\begin{equation}
    \frac{d}{dt}
    \begin{pmatrix}
    \tilde{\alpha}_k^*\\ 
    \tilde{\alpha}_{-k}
    \end{pmatrix}
    =
    \mathrm{i}
    \begin{pmatrix}
    \tilde{\epsilon}_k-\mathrm{i}\frac{\gamma\mathbb{I}+\eta\left|\alpha_k\right|^2}{2} & 
    \frac{\left[\tilde{H}_c\right]_{12}}{2} \\
    -\frac{\left[\tilde{H}_c\right]_{21}}{2} &
    -\tilde{\epsilon}_{-k}-\mathrm{i}\frac{\gamma\mathbb{I}+\eta\left|\alpha_k\right|^2}{2}
    \end{pmatrix}
    \begin{pmatrix}
    \tilde{\alpha}_k^*\\ 
    \tilde{\alpha}_{-k}
    \end{pmatrix},
    \label{eq::EOM_Appendix}
\end{equation}
}where $\hat{\tilde{a}}_{n,k}$ is the magnon annihilation operator of $n$-th band at $k$-point; $\tilde{\epsilon}_{k}$ is a diagonal matrix with elements $\epsilon_{n,k}-\frac{\Omega}{2}$, where $\epsilon_{n,k}$ is the energy eigenvalue;
%is the diagonal matrix of eigenvalues of matrix $H_0(k)$ subtracted by $\frac{\Omega}{2}$; 
$\gamma$ and $\eta$ are phenomenological linear and non-linear damping constants; 
$\mathbb{I}$ is identity matrix and $\left|\alpha_k\right|^2$ is diagonal matrix with entries $\left|\left\langle\hat{\tilde{a}}_{n,k}\right\rangle\right|^2$.

%\section{Quartic Interaction Term}

%The four body interaction term that is used for the calculation of two magnon scattering rate is derived from the spin Hamiltonian using higher order terms of Taylor-series expansion of square root in Holstein Primakoff transformation. The interaction term is given by,
%\begin{equation}
%    \pazocal{H}_{0,\text{int}}^{(4)}=-\frac{J\hbar^2}{4} \sum_{\left\langle ij\right\rangle}
%    \left[ 
%    4\hat{a}_i^\dagger\hat{a}_j^\dagger\hat{a}_i\hat{a}_j
%    +\hat{a}_j^\dagger \hat{a}_j^\dagger \hat{a}_i\hat{a}_j
%    +\hat{a}_i^\dagger \hat{a}_j^\dagger \hat{a}_i\hat{a}_i
%    +\hat{a}_i^\dagger\hat{a}_j^\dagger \hat{a}_j\hat{a}_j
   % +\hat{a}_i^\dagger\hat{a}_i^\dagger\hat{a}_i\hat{a}_j
 %   \right]
%\end{equation}

\section{\label{Appendix::C}Equilibration of spin oscillation}
In this section, we show that the oscillation at different $k$-points equillibrate at different times (see Fig.\,\ref{fig::TimeCrystalAppendix}(a),(b)) resulting in a transient modulation of amplitude before reaching the steady state as in Fig.\,\ref{fig::TimeCrystal}(d) in main text.
The figure also shows that the amplitude of oscillation decreases rapidly away from $k=\pi$.
\begin{figure}[h]
\includegraphics[width=0.8\textwidth]{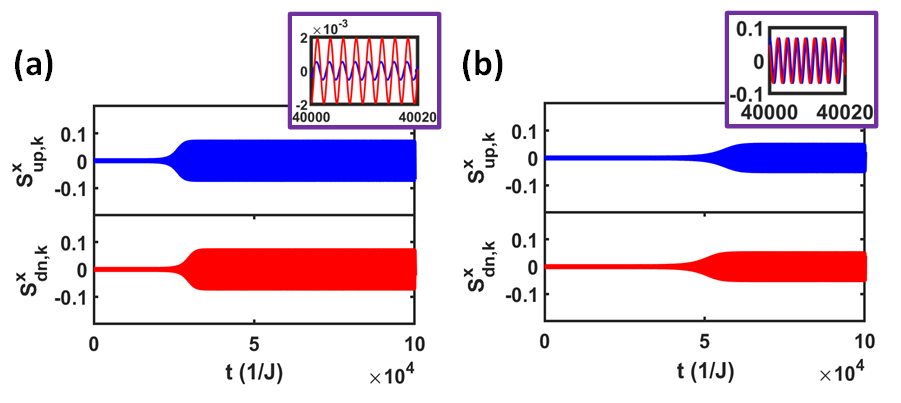} % this command will be ignored
\caption{The oscillation of spin component $S^x$ at a site at upper edge (blue plot) and lower edge (red plot) are calculated (d) at $k=\pi$, (e) at $k=3.1793$ in Brillouin zone for a system with $N_y=91$. Insets show the magnified figures of the corresponding figures within a particular time limit. The parameters used for all the plots are $J=1.0$, $B_z\rightarrow 0^+$, $\gamma=5\times 10^{-4}$, $\eta=9\times 10^{-4}$, $\Omega=5.1716$, $p_0=1.0$, $E_0^x=0.0$, $E_0^y=0.002$. The figure shows that the different $k$-points arrive at steady state at a different time and the amplitude of oscillation decreases rapidly away from $k=\pi$.}
\label{fig::TimeCrystalAppendix}
\end{figure}

\section{\label{Appendix::D}A toy-model for analytical calculation of dynamics}

In the main text we showed that the time-crystal is not stable in absence of effective time reversal symmetry.
The reason behind this chaotic oscillation can be understood from the following simple model with only two coupled modes with energies $\omega_k$ and $\omega_{-k}$,
\begin{equation}
    \pazocal{H}=\omega_k \hat{a}_k^\dagger \hat{a}_k + \omega_{-k} \hat{a}_{-k}^\dagger \hat{a}_{-k} +2\epsilon \cos(\Omega t) \left(\hat{a}_k^\dagger \hat{a}_{-k}^\dagger + \hat{a}_k \hat{a}_{-k}\right)
\end{equation}
Next we rotate the basis states with the unitary operator $U=e^{i\Omega(\hat{n}_k+\hat{n}_{-k})t}$, where $\hat{n}_k=\hat{a}_k^\dagger \hat{a}_k$.
The Hamiltonian in rotating frame is,
\begin{align}
 \pazocal{H}^\prime &= U(t) \pazocal{H} U^\dagger(t) + i\hbar \Dot{U}(t) U^\dagger(t) \nonumber\\
 &=\tilde{\omega}_k \hat{a}_k^\dagger\hat{a}_k
 +\tilde{\omega}_{-k} \hat{a}_{-k}^\dagger\hat{a}_{-k}
 +\epsilon \left(e^{2i\Omega t}+1\right)\hat{a}_k^\dagger \hat{a}_{-k}^\dagger
 +\epsilon \left(1+e^{-2i\Omega t}\right)\hat{a}_k \hat{a}_{-k},
\end{align}
where $\tilde{omega}_k=\omega_k-\Omega/2$. Neglecting fast rotations (terms with $e^{\pm 2i\Omega t}$), the effective Hamiltonian in rotating wave approximation becomes,
\begin{align}
    \pazocal{H}_{\text{eff}}=\tilde{\omega}_k \hat{a}_k^\dagger\hat{a}_k
 +\tilde{\omega}_{-k} \hat{a}_{-k}^\dagger\hat{a}_{-k}
 +\epsilon \hat{a}_k^\dagger \hat{a}_{-k}^\dagger
 +\epsilon \hat{a}_k \hat{a}_{-k}
\end{align}
By defining the field $\alpha_k=\left\langle \hat{a}_k\right\rangle$ and using the equation of motion $i\frac{d}{dt}\left\langle \hat{O}\right\rangle=\left\langle\left[\hat{O}, \hat{H}_{\text{eff}}\right]\right\rangle$, we get the following coupled differential equation,
\begin{equation}
    i\frac{d}{dt}
    \begin{pmatrix}
    \alpha_k\\
    \alpha_{-k}^*
    \end{pmatrix}
    =
    \begin{pmatrix}
    \tilde{\omega}_k-i\frac{\gamma}{2} & \epsilon \\
    -\epsilon & -\tilde{\omega}_{-k}-i\frac{\gamma}{2} 
    \end{pmatrix}
    \begin{pmatrix}
    \alpha_k \\
    \alpha^*_{-k}
    \end{pmatrix},
    \label{eq::DifferentialEquation}
\end{equation}
where the phenomenological damping $\gamma$ is added.
The coupled differential equation can be transformed into the following second order differential equation,
\begin{equation}
    \frac{d^2\alpha_k}{dt^2}-\left(\tilde{\omega}_k-\tilde{\omega}_{-k}-i\gamma\right)\frac{d\alpha_k}{dt}+\left[\epsilon^2-\left(\tilde{\omega}_k-i\frac{\gamma}{2}\right)\left(\tilde{\omega}_{-k}+i\frac{\gamma}{2}\right)\right]=0
\end{equation}
Now using the ansatz $\alpha_k=A_1 e^{-i\omega t}$ we get an quadratic equation in $\omega$ and the solution of which is given by,
\begin{equation}
    \omega=\frac{\omega_k-\omega_{-k}}{2}
    -i\frac{\gamma}{2}
    \pm i\sqrt{\epsilon-\frac{(\tilde{\omega}_k+\tilde{\omega}_{-k})^2}{4}}
\end{equation}
The solution with positive sign is the only physical solution, because in the limit $\epsilon\rightarrow 0$ the frequency should be $\omega\rightarrow\omega_k$. Thus the solution for the field,
\begin{equation}
    \alpha_k=A_+ e^{-i\frac{(\omega_k-\omega_k) t}{2}} 
    e^{-\frac{\gamma t}{2}} 
    e^{\sqrt{\epsilon-\frac{(\tilde{\omega}_k+\tilde{\omega}_{-k})}{4}}t},
\end{equation}
Thus the condition for amplification is given by,
\begin{equation}
    \sqrt{\epsilon-\frac{(\tilde{\omega}_k+\tilde{\omega}_{-k})}{4}}>\frac{\gamma}{2}.
\end{equation}
From now on, we focus on the oscillatory part of the solution, by taking the exponential decay or amplification into the amplitude,
\begin{equation}
    \alpha_k=A_+(t) e^{-i\frac{(\omega_k-\omega_{-k}) t}{2}}
\end{equation}
From the Eq.\ref{eq::DifferentialEquation}, we get,
$\alpha_{-k}^*=A_{-}(t) e^{-i\frac{(\omega_k-\omega_{-k}) t}{2}}$, and thus we have,
\begin{equation}
    \alpha_{-k}=A_{-}(t) e^{i\frac{(\omega_k-\omega_{-k}) t}{2}}
\end{equation}

It can be shown that the relation between the fields in lab and rotating frame is given by $\alpha_{\pm k}^{\text{lab}}=e^{-i\frac{\Omega}{2}t}\alpha_{\pm k}$, thus the fields in lab frame is given by,
\begin{equation}
    \alpha_{\pm k}^{\text{lab}}=A_{\pm}(t) e^{\mp i\frac{(\omega_k-\omega_{-k})t}{2}}
    e^{-i\frac{\Omega}{2}t}.
\end{equation}
From this equation, we can conclude the oscillation frequency due to parametric amplification is half of the driving field only when the coupled state which is amplified have equal energies $\omega_k=\omega_{-k}$, otherwise the frequency of oscillation is different.
Moreover, when effects of oscillation from many coupled oscillators with $\omega_k\neq \omega_{-k}$ are superimposed as in equation Eq.\,\ref{eq::Sx} in main text, then the resultant oscillation will be chaotic, which is visible in Fig.\,\ref{fig::TimeCrystal2}(b).

\begin{figure}[t]
\includegraphics[width=0.7\textwidth]{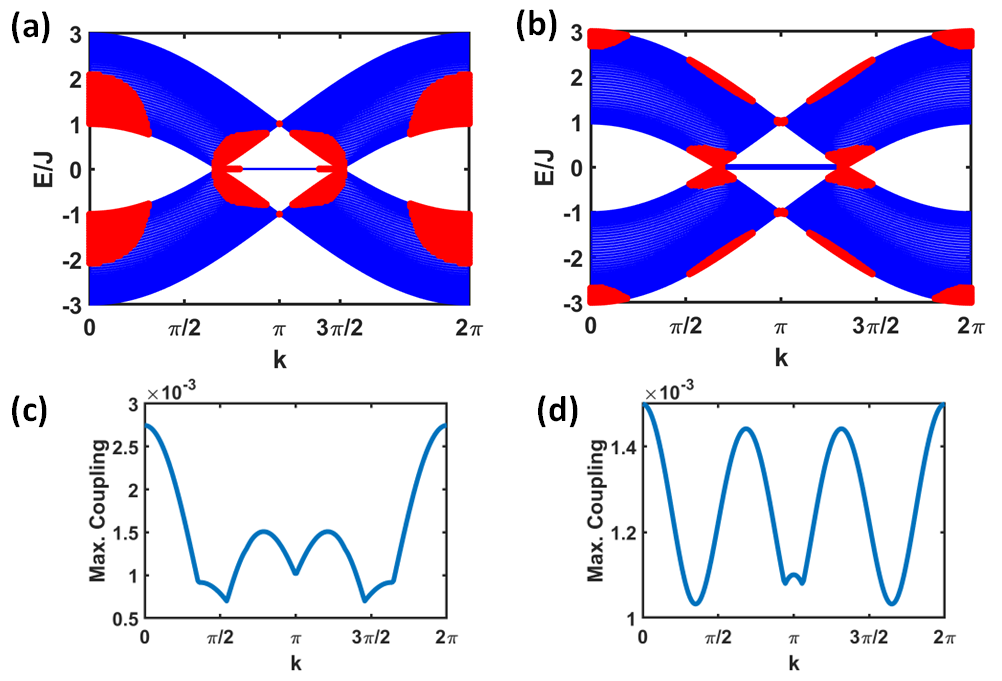} % this command will be ignored
\caption{The magnon band structure is shown in blue, whereas the red dots denote eigenstate with eigenvalues with positive imaginary part for electric fields (a) $E_0^y=0.015$ and (b) $E_0^x=0.015$. Maximum of absolute values of matrix elements $\left[\tilde{H}_c\right]_{12}$ is ploted as a function of momentum for electric field amplitudes (c) $E_0^y=0.001$ and (d) $E_0^x=0.002$. The other parameters used for all the plots are $J=1.0$, $S=1$, $B_z\rightarrow 0^+$, $\gamma=2.5\times 10^{-3}$, $\Omega=3JS$, $K^{x,xy}=K^{y,xx}=K^{y,yy}=1.0$.}
\label{fig::Honeycomb}
\end{figure}

\section{\label{Appendix::E}Comparision between Kagome lattice and Honeycomb lattice}
In this section we show that Kagome lattice structure is a suitable lattice structure for achieving magnon time crystal, by comparing the results with honeycomb lattice.
We have studied a similar model in honeycomb lattice to show the advantage of the Kagome lattice structure.
The model Hamiltonian we take on Honeycomb lattice is given as,
\begin{align}
    \pazocal{H}=-J\sum_{\left\langle ij\right\rangle} \hat{\boldsymbol{S}}_i\cdot\hat{\boldsymbol{S}}_j + E^x(t) \sum_i K_i^{x,xy} \left[\hat{S}^x_{j,A} \hat{S}^y_{j,A}-\hat{S}^x_{j,B} \hat{S}^y_{j,B}\right]
    +E^y(t)\sum_i &\left[K_j^{y,xx}\left(\hat{\boldsymbol{S}}^x_{j,A}\hat{\boldsymbol{S}}^x_{j,A}+\hat{\boldsymbol{S}}^x_{j,B}\hat{\boldsymbol{S}}^x_{j,B}\right)\right.
    \nonumber\\
    &+\left. K_j^{y,yy} \left(\hat{\boldsymbol{S}}^y_{j,A}\hat{\boldsymbol{S}}^y_{j,A}+\hat{\boldsymbol{S}}^y_{j,B}\hat{\boldsymbol{S}}^y_{j,B}\right)\right],
\end{align}
where the polarization terms are considered respects the symmetry of the lattice and the terms which does not contribute to the amplification is already discarded (mathematically those terms will be discarded in rotating wave approximation).
By diagonalizing the dynamical matrix as discussed in the main text  we achieved the band structure as shown in the Fig.\,\ref{fig::Honeycomb}(a) and (b) for different polarization.
It can be noticed that the electromagnetic field amplifies the bulk magnon bands instead of edge magnon bands, which is backed up by the results in Fig.\,\ref{fig::Honeycomb}(c) and (d) showing maximum coupling with electromagnetic field occurs with the bulk magnon states.

Whereas, the Kagome lattice structure is useful for amplification of edge magnons without amplifying the bulk magnons.
The reason behind not amplifying the bulk magnons is for making the system more stable. Fig.\,\ref{fig::TimeCrystal3}(a) and (b) shows that the magnon amplication in the Kagome lattice is confined to the edge states only.
Moreover, choice of polarization of electromagnetic field should be in y-direction to specifically amplify the edge magnons.
The Fig.\ref{fig::TimeCrystal3}(a) and (b) shows that the imaginary parts of bulk magnon modes zero and non-zero for y-polarized and x-polarized electromagnetic field.
The reason of this discrepancy is due to the difference in coupling terms as shown in Fig.\,\ref{fig::TimeCrystal3}(c) and (d).
The y-polarized EM-field strongly couples with edge magnons, whereas the x-polarized EM-field strongly couples with bulk magnons.

\begin{figure}[t]
\includegraphics[width=0.65\textwidth]{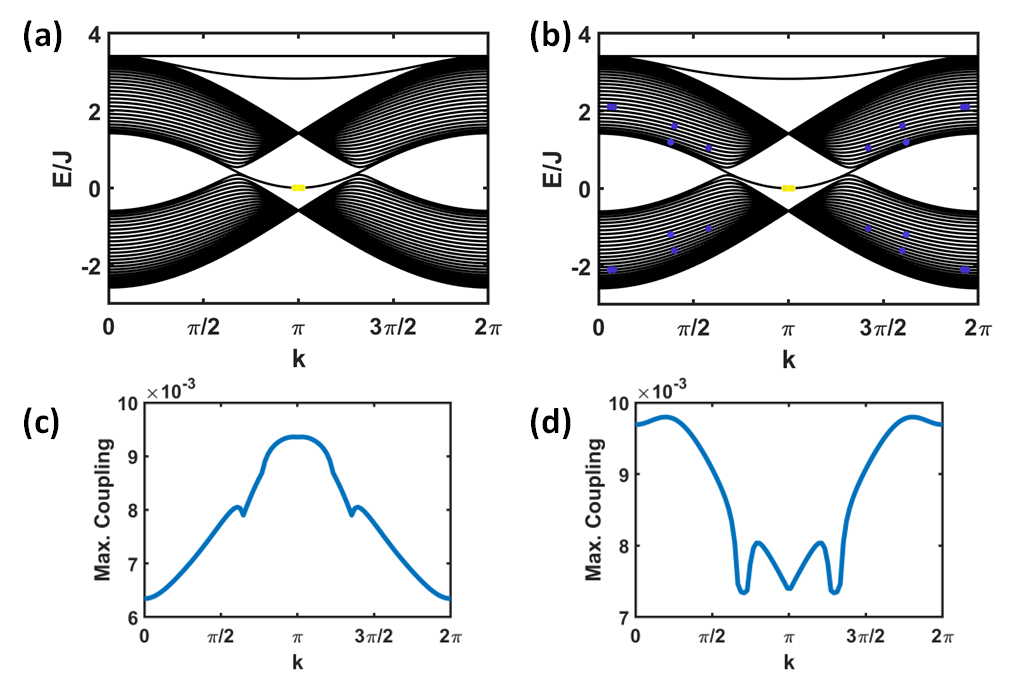} % this command will be ignored
\caption{The magnon band structure is shown in black; the yellow and blue dots denote eigenstate with eigenvalues with positive imaginary part for electric fields (a) $E_0^y=0.015$ and (b) $E_0^x=0.015$. Maximum of absolute values of matrix elements $\left[\tilde{H}_c\right]_{12}$ is ploted as a function of momentum for electric field amplitudes (c) $E_0^y=0.015$ and (d) $E_0^x=0.015$. The other parameters used for all the plots are $J=1.0$, $D=0.00$,  $B_z\rightarrow 0^+$, $\gamma=5\times 10^{-4}$, $\Omega=5.1716$, $p_0=1.0$.}
\label{fig::TimeCrystal3}
\end{figure}

\section{\label{Appendix::F}Required intensity of light and the effect of magnetic field of electromagnetic wave}
In this section, we discuss about the effect of magnetic field and intensity of light that is required for amplification procedure.
The intensity of light that is required to get amplification is unknown and depends on damping of magnon states. 
The lifetime of magnons can be $1\mu s$, $1ns$, $1ps$ etc. Depending on that the energy-state broadening according to uncertainty principle would be $10^{-9}eV$, $10^{-6}eV$,$10^{-3}eV$ respectively.
Thus, the damping parameters should be  $\lambda=10^{-9}eV$, $10^{-6}eV$,  $10^{-3}eV$ respectively.
Now to have amplification, one requires to have electric field amplitude $E_c$ (V/nm) which should follow the inequality relationship for a system with damping $\lambda$, (considering the following ideal system parameters, lattice constant $a$=1nm, $t/U=10^{-2}$)

\begin{align}
E_c P &\geq \lambda [\text{eV}] \nonumber\\
E_c a [\text{nm}] e (t/U)^3 &\geq \lambda [\text{eV}]\:\left[\text{ where, }P\approx ae\frac{t^3}{U^3}\right] \nonumber\\
E_c \times 10^{-6} &\geq \lambda \,\text{V/nm} \nonumber\\
E_c &\geq 10^6\lambda \,\text{V/nm} \nonumber\\
E_c &\geq 10^{15}\lambda \,\text{V/m}
\end{align}
Thus the electric field required for the amplification should be $10^6V/m$, $10^9V/m$, $10^{12}V/m$ respectively.
Based on the relation B=E/c , the magnetic field amplitude should be, B = 0.01 T, 1 T, 1000 T respectively. 
The magnetic  field 0.01T and 1T are still very negligible for a magnetic insulator, because the Heisenberg exchange interaction is $1meV$ whereas the energy equivalent to 1T is $\mu_B B=0.01meV$.

\end{widetext}

%\bibliographystyle{apsrev4-PRX}
%\bibliography{ref}

\end{document}